\documentclass[11pt,english]{article}

\usepackage{hyperref,amsfonts,amsmath,amssymb,bm,a4wide,cite,graphicx,babel}

\usepackage[T1]{fontenc}
\usepackage[latin9]{inputenc}

\numberwithin{equation}{section}

\begin{document}

\title{\textbf{Cherenkov plasmons emission by primordial neutrinos}}

\author{Maxim Dvornikov\thanks{maxim.dvornikov@gmail.com}
\\
\small{\ Pushkov Institute of Terrestrial Magnetism, Ionosphere} \\
\small{and Radiowave Propagation (IZMIRAN),} \\
\small{108840 Moscow, Troitsk, Russia}}

\date{}

\maketitle

\begin{abstract}
We study the emission of Cherenkov plasmons by the gas of neutrinos
with a nonzero temperature and a chemical potential. The background plasma,
consisting of charged leptons, is taken to be nonrelativistic. The
energy emission rate is obtained for longitudinal plasmons. To get
the neutrino emissivity we average quantum field theory matrix element
over the distribution functions of incoming and outgoing particles.
Our results are applied for the description of the cooling down of
a neutrino cluster formed in the early universe. Such clusters can exist
owing to the neutrino interaction with a hypothetical light scalar
boson. Using particular cluster parameters, we demonstrate that the
considered cooling mechanism is efficient for some clusters. We find
the temperature range where the proposed cooling channel is valid.
Some useful calculations of the polarization tensor, as well as the
plasmon form factors and their dispersion relations are also provided.
\end{abstract}

\section{Introduction}

A great fraction of the mass in our universe was established in numerous
astronomical observations to consist of dark matter. The essence of
dark matter is unclear. One can just say that it is nonbaryonic~\cite[pp.~159--173]{Wei20}.
Multiple candidates are considered to play the role of dark matter.
First, we mention weakly interacting massive particles~\cite{RosSesTro18}.
Nowadays, light pseudoscalar particles, called axions and axion like
particles (ALPs), are assumed to be the most plausible candidates
for dark matter constituents~\cite{Kim87}. Nevertheless, besides
the claims in Ref.~\cite{Ber13}, no signals for dark matter particles
have been detected in laboratory experiments yet (see, e.g., Ref.~\cite{Boz25}).

Previously, massive neutrinos were thought to contribute to dark matter~\cite{Gun78}.
However, later, it was understood (see, e.g., Ref.~\cite[pp.~190--191]{Wei20})
that this kind of neutrinos cannot be cold dark matter since they
are relativistic. Nevertheless, if we assume that light massive neutrinos
can form a condensate around galaxies, these particles do not diffuse
and can contribute to dark matter to some extent. It is clear that
the formation of a neutrino condensate implies an attractive interaction
beyond the standard model. The mediator of this interaction can be
a hypothetical light scalar particle weakly coupled to neutrinos.
Thus, such a scalar boson can be a competitor to axions and ALPs for
a dark matter constituent. The recent development of the neutrino condensate contribution to dark matter is given in Ref.~\cite{Cli06}.

The formation of neutrino clusters owing to a scalar boson interaction
was first considered in Ref~\cite{Ste98}. Then, the formation of
neutrino clusters was studied in Ref.~\cite{SmiXu22}, where multiple
masses of a scalar boson and coupling constants to neutrinos were considered. Besides the formation of a cluster, the interaction
with a scalar particle makes the neutrino condensate to be superfluid~\cite{Kap04}.
The neutrino superfluidity owing to the scalar boson interaction was
studied in Refs.~\cite{Aza11,Dvo24}. The implication of the neutrino
condensation to the dark matter problem was considered in Ref.~\cite{Add22}.
Recently, the formation of dark matter halos in the presence of the
neutrino condensate was studied in Ref.~\cite{Cap26}.

When a cluster appears owing to an instability, the neutrino gas is
compressed and its temperature increases. Thus, a cluster can decay
because of the thermal motion of neutrinos. Moreover, if one expects
the superfluid neutrino condensate in such a cluster, additional thermal
fluctuations can destroy the supefluidity. The problem of the cluster
cooling was raised in Ref.~\cite{SmiXu22}. However, the mechanisms
to carry away the thermal energy from a cluster, proposed in Ref.~\cite{SmiXu22},
are not efficient.

In Refs.~\cite{Dvo24,Dvo25}, we put forward the alternative cooling
mechanism based on the Cherenkov plasmons emission. We assumed that
a neutrino cluster is formed in the early universe. Then, accounting
for the parameters of a cluster from numerical simulations, we estimated
the cooling time. Requesting that the cooling time is less than the
universe age, we obtained the range of the primordial plasma temperature, or the universe age, favorable for the cluster
formation.

It is known that the Cherenkov radiation is allowed for charged particles
whereas a neutrino is a neutral fermion. Nevertheless, it can acquire
the induced electric charge while propagating in background medium~\cite{OraSemSmo94}.
The Cherenkov radiation was shown in Ref.~\cite{OliNiePal96} to exist
even for standard model massless neutrinos. This kind of the radiation
by supernova (SN) neutrinos was studied in Ref.~\cite{Sah97}.

Despite we demonstrated in Refs.~\cite{Dvo24,Dvo25} that a neutrino
cluster can potentially cool down by the Cherenkov plasmons emission,
the consideration of the cooling process in Refs.~\cite{Dvo24,Dvo25}
had some shortcomings. The description of the energy emission in Ref.~\cite{Dvo24}
was quite sketchy. In Ref.~\cite{Dvo25}, we adopted the ultrarelativistic
plasma model. However, the temperature interval for the cluster formation
was extrapolated to the nonrelativistic range. The aim of the present
work is the detailed study of the cluster cooling in case of the nonrelativistic
background plasma which is more interesting from the point of view
of phenomenological applications.

This work is organized in the following way. First, in Sec.~\ref{sec:CHEREMGEN},
we derive the matrix element for the Cherenkov radiation and the emissivity
of the neutrino gas. Then, in Sec.~\ref{sec:CLUSTER}, we apply our
results for the description of the particular cluster cooling. Finally,
we conclude in Sec.~\ref{sec:CONCL}. Some useful expressions from
the finite temperature field theory are provided in Appendix~\ref{sec:SUMFTFT}.
The properties of plasmons in background medium are described in Appendix~\ref{sec:PLASMPROP}.
We calculate the form factors of plasmons in Appendix~\ref{sec:CALCPLASMFF}.
The dispersion relations for plasmons in nonrelativistic medium are
established in Appendix~\ref{sec:DISPREL}.

\section{Cherenkov emission by a neutrino gas}\label{sec:CHEREMGEN}

In this section, we derive the emissivity of the neutrino gas with
the nonzero temperature and the chemical potential.

\begin{figure}
  \centering
  \includegraphics[scale=1]{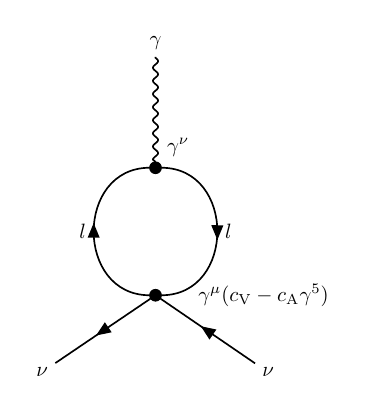}
  \protect
  \caption{The Feynman diagram for the neutrino Cherenkov emission $\nu\to\nu+\gamma$
which results in the matrix element in Eq.~(\ref{eq:matreldef}).\label{fig:feyndiag}}
\end{figure}

The matrix element for the process $\nu\to\nu+\gamma$, with the Feynman
diagram being depicted in Fig.~\ref{fig:feyndiag}, reads
\begin{equation}\label{eq:matreldef}
  \mathcal{M}=\frac{eG_{\mathrm{F}}}{\sqrt{2}}
  \bar{u}_{2}\gamma_{\mu}(1-\gamma^{5})
  u_{1}\tilde{\Pi}^{\mu\nu}e_{\nu}(k),
\end{equation}
where $u_{1,2}=u(p_{1,2})$ are the bispinors of the incoming and
the outgoing neutrinos having the four momenta $p_{1,2}^{\mu}=(E_{1,2},\mathbf{p}_{1,2})$,
$\gamma^{\mu}=(\gamma^{0},\bm{\gamma})$ and $\gamma^{5}=\mathrm{i}\gamma^{0}\gamma^{1}\gamma^{2}\gamma^{3}$
are the Dirac matrices, $e_{\mu}(q)$ is
the plasmon polarization vector with $k^\mu=p_{1}^\mu-p_{2}^\mu$, $e$ is the
elementary charge, $G_{\mathrm{F}}=1.17\times10^{-5}\,\text{GeV}^{-2}$
is the Fermi constant, and
\begin{equation}\label{eq:Pimunudef}
  \tilde{\Pi}^{\mu\nu}(k)=-\int\frac{\mathrm{d}^{4}P}{(2\pi)^{4}}\frac{\mathrm{tr}
  \left[
    (\not P+m)\gamma^{\mu}(c_{\mathrm{V}}-c_{\mathrm{A}}\gamma^{5})(\not P-\not k+m)\gamma^{\nu}
  \right]}
  {[P^{2}-m^{2}][(P-k)^{2}-m^{2}]},
\end{equation}
is the generalized polarization tensor of a plasmon. The image for
$\tilde{\Pi}^{\mu\nu}$ can be obtained from Fig.~\ref{fig:feyndiag} by
the truncation of the external lines. In Eq.~(\ref{eq:Pimunudef}),
$c_{\mathrm{V,A}}$ are the vector and axial vector coupling constants,
and $m$ is the mass of a charged lepton $l$ counterpart to $\nu$.

The Cherenkov emission is forbidden in vacuum, where the photon dispersion
is $k^{2}=0$. However, in matter, where $k^{2}\neq0$ for a plasmon,
this process is allowed~\cite{OliNiePal96}. In this situation, we
have to take into account the nontrivial plasmon dispersion and evaluate $\tilde{\Pi}^{\mu\nu}$
in Eq.~(\ref{eq:Pimunudef}) in matter where the temperature and the chemical potential are nonzero.

The plasmon properties in medium can be inferred from the plasmon polarization tensor $\Pi_{\mu\nu}$, which is related to $\tilde{\Pi}_{\mu\nu}$ in Eq.~(\ref{eq:Pimunudef}) by $\Pi_{\mu\nu} = e^2 \tilde{\Pi}_{\mu\nu}/c_\mathrm{V}$,
and $c_\mathrm{A} = 0$. We calculate $\Pi_{\mu\nu}$ using the imaginary
time perturbation theory. Some basic integrals are computed in Appendix~\ref{sec:SUMFTFT}.
We calculate the plasmon form factors, resulting from $\Pi_{\mu\nu}$, in
Appendix~\ref{sec:CALCPLASMFF}. Following Ref.~\cite{NiePal89}, we represent $\Pi_{\mu\nu}$ in a nonchiral medium in the form,
\begin{equation}\label{eq:Pimunudecom}
  \Pi_{\mu\nu}=\Pi_{\mathrm{L}}Q_{\mu\nu}+\Pi_{\mathrm{T}}R_{\mu\nu},
\end{equation}
where $\Pi_{\mathrm{L,R}}$ are the longitudinal and transverse form factors. The projection
operators in Eq.~(\ref{eq:Pimunudecom}), $Q_{\mu\nu}$ and $R_{\mu\nu}$, are given in Eq.~\eqref{eq:projoperdef}. Note that the tensor $\tilde{\Pi}_{\mu\nu}$ can be also decomposed analogously to Eq.~\eqref{eq:Pimunudecom}.

Squaring the matrix element in Eq.~(\ref{eq:matreldef}), one gets
that
\begin{equation}\label{eq:M2}
  |\mathcal{M}|^{2}=\frac{e^{2}G_{\mathrm{F}}^{2}}{2}
  \mathrm{tr}[\rho_{1}\gamma_{\alpha}(1-\gamma^{5})\rho_{2}\gamma_{\beta}(1-\gamma^{5})]
  \tilde{\Pi}^{*\alpha\mu}e_{\mu}^{*}(k)\tilde{\Pi}^{\beta\nu}e_{\nu}(k),
\end{equation}
where $\rho_{1,2}=u_{1,2}\bar{u}_{1,2}=(\gamma_{\lambda}p_{1,2}^{\lambda})(1+\gamma^{5})/2$
are the spin density matrices of ultrarelativistic neutrinos~\cite[p.~115]{BerLifPit82}.

We should sum $|\mathcal{M}|^{2}$ in Eq.~(\ref{eq:M2}) over the
polarizations $\lambda$ of a plasmon. In medium with a nonzero temperature and chemical potential, we have one longitudinal, $\lambda=\mathrm{L}$, and two transverse plasmons, $\lambda\equiv s=1,2$. The standard model plasma consisting of leptons and neutrinos, which Eq.~\eqref{eq:Pimunudef} corresponds to, is nonchiral. In this situation, two transverse plasmons are degenerate; cf. Appendix~\ref{sec:PLASMPROP}. Therefore,
\begin{align}\label{eq:M2sumpol}
  \sum_{\lambda}|\mathcal{M}|^{2}= & 4e^{2}G_{\mathrm{F}}^{2}
  [p_{1}^{\mu}p_{2}^{\nu}+p_{2}^{\mu}p_{1}^{\nu}-g^{\mu\nu}(p_{1}p_{2})]
  \left(
    \tilde{\Pi}_{\mathrm{L}}^{2}Q_{\mu\nu}-\tilde{\Pi}_{\mathrm{T}}^{2}R_{\mu\nu}
  \right)
  \nonumber
  \\
  & =
  \frac{4 c_\mathrm{V}^2 G_{\mathrm{F}}^{2}}{e^2}
  \bigg\{
    \Pi_{\mathrm{L}}^{2}(\omega,K)
    \left(
      1-\frac{\omega^{2}}{K^{2}}
    \right)
    \left[
      2E_{1}(E_{1}-\omega)+\frac{\omega^{2}-K^{2}}{2}
    \right]_{\mathrm{long}}
    \nonumber
    \\
    & +
    \Pi_{\mathrm{T}}^{2}(\omega,K)
    \left(
      1-\frac{\omega^{2}}{K^{2}}
    \right)
    \left[
      2E_{1}(E_{1}-\omega)+\frac{\omega^{2}+K^{2}}{2}
    \right]_{\mathrm{trans}}
  \bigg\},
\end{align}
where the subscripts `long' and `trans' mean that one should account for either longitudinal
or transverse dispersion relations which are given in Eqs.~(\ref{eq:disprelLfin})
and~(\ref{eq:disprelT}). 

To derive Eq.~(\ref{eq:M2sumpol}) we
represent the plasmon momentum as $k^\mu=(\omega,\mathbf{K})$.
Additionally, we use the conservation laws,
\begin{equation}\label{eq:4momcons}
  E_{1}=E_{2}+\omega,\quad\mathbf{p}_{1}=\mathbf{p}_{2}+\mathbf{K},
\end{equation}
with both neutrinos and the plasmon being on mass shell. The longitudinal and transverse form factors, $\Pi_{\mathrm{L,R}}$ are  defined in Eq.~\eqref{eq:Pimunudecom}. The quantities $\tilde{\Pi}_{\mathrm{L,R}}$  and $\Pi_{\mathrm{L,R}}$ are related by $\tilde{\Pi}_{\mathrm{L,R}} = c_\mathrm{V}\Pi_{\mathrm{L,R}}/e^2$. It happens since, at $c_\mathrm{A} = 0$, $\tilde{\Pi}_{\mu\nu}$ coincides with $\Pi_{\mu\nu}$ up to a constant factor. That is why, we use the same projection operators $Q_{\mu\nu}$ and $R_{\mu\nu}$, defined in Eq.~(\ref{eq:projoperdef}), to decompose $\tilde{\Pi}_{\mu\nu}$ in Eq.~\eqref{eq:Pimunudef}. We mentioned this fact earlier. Moreover,
in Eq.~\eqref{eq:M2sumpol}, we take into account the properties of $Q_{\mu\nu}$ and $R_{\mu\nu}$ given in Eq.~(\ref{eq:projoperorth}).

Based on Eq.~(\ref{eq:4momcons}), one gets that the Cherenkov emission
is possible if $K>\omega$. Using Eq.~(\ref{eq:disprelLfin}),
we obtain that, for longitudinal plasmons, this condition is fulfilled
when $\omega>\omega_{p}\sqrt{1+3T/m}$, where $\omega_{p}$ is the
plasma frequency. On the contrary, based on Eq.~(\ref{eq:disprelT}),
we get that the condition $K>\omega$ is never satisfied for transverse
plasmons. Analogous result was obtained in Ref.~\cite{Sah97} for
an ultrarelativistic plasma. Therefore, we take into account only
the longitudinal plasmons contribution in Eq.~(\ref{eq:M2sumpol}).

Suppose that the neutrino gas occupies the volume $V$. The emissivity
of the gas by Cherenkov plasmons, i.e. the energy emitted per unit
time, reads
\begin{align}\label{eq:dotEdef}
  \dot{E}= & \int\frac{V\mathrm{d}^{3}p_{1}}{2E_{1}(2\pi)^{3}}
  \frac{\mathrm{d}^{3}p_{2}}{2E_{2}(2\pi)^{3}}
  \frac{\mathrm{d}^{3}K}{2\omega(2\pi)^{3}}\omega(2\pi)^{4}\delta^{4}(p_{2}+k-p_{1})
  \nonumber
  \\
  & \times\frac{\omega}{K}\frac{\mathrm{d}\omega}{\mathrm{d}K}\sum_{\lambda}|\mathcal{M}|^{2}f_{1}(1-f_{2}),
\end{align}
where $f_{1,2}=\left\{ \exp[(E_{1,2}-\mu_{\nu})/T_{\nu}]+1\right\} ^{-1}$
are the distribution functions of incoming and outgoing neutrinos,
corresponding to the temperature $T_{\nu}$ and the chemical potential
$\mu_{\nu}$. Shortly in Sec.~\ref{sec:CLUSTER}, we apply our results for the cluster cooling.
In this situation, the neutrino temperature inside a cluster is not equal to that of
charged leptons, $T_{\nu}\equiv T_{\mathrm{clust}}\neq T$.
The matrix element, summed over the plasmon polarizations (in fact we have only
one polarization for a longitudinal plasmon), is given in Eq.~(\ref{eq:M2sumpol}).
Finally, we mention that, in Eq.~(\ref{eq:dotEdef}), we account
for the electric charge renormalization in Eq.~(\ref{eq:renorm}).

The integration over the momentum of an outgoing neutrino is made
with help of the momentum conservation $\delta$-function~\cite{Sah97}.
Then, we integrate over the angle between $\mathbf{p}_{1}$ and $\mathbf{K}$
using the remaining energy conservation $\delta$-function~\cite{Sah97}.
Changing the integration variable $K\to\omega$ (see, Appendix~\ref{sec:PLASMPROP}),
one obtains the range of the $\omega$ variation, $\omega_{\mathrm{min}}<\omega<\omega_{\mathrm{max}}$.
The maximal value $\omega_{\mathrm{max}}$ is the solution of the
equation
\begin{equation}\label{eq:omegamaxeq}
  K(\omega)+\omega=2E_{1},
\end{equation}
where $K(\omega)$ is the dispersion relation for longitudinal plasmons
in Eq.~\eqref{eq:disprelLfin}. The minimal value $\omega_{\mathrm{min}}$
is again given by Eq.~(\ref{eq:disprelLfin}) at $K=\omega$. It
is interesting to mention that the condition $\omega<\omega_{\mathrm{max}}$,
with $\omega_{\mathrm{max}}$ given by Eq.~(\ref{eq:omegamaxeq}),
guarantees that the matrix element in Eq.~(\ref{eq:M2sumpol}) is
positive.

Using the fact that $\omega^{2}-K^{2}=\Pi_{\mathrm{L}}(\omega,K)$
for longitudinal plasmons, we rewrite Eq.~(\ref{eq:dotEdef}) in
the form,
\begin{align}\label{eq:dotEfin}
\dot{E}= & \frac{VG_{\mathrm{F}}^{2}c_{\mathrm{V}}^{2}}{8\pi^{3}e^{2}}\int_{0}^{\infty}\mathrm{d}E_{1}\int_{\omega_{\mathrm{min}}}^{\omega_{\mathrm{max}}}\omega^{5}\mathrm{d}\omega f(E_{1})\left[1-f(|E_{1}-\omega|)\right]\left(1-\frac{\omega^{2}}{K^{2}}\right)^{3}\nonumber \\
 & \times\left[2E_{1}(E_{1}-\omega)+\frac{1}{2}\left(\omega^{2}-k^{2}\right)\right],
\end{align}
where $c_{\mathrm{V}}=\tfrac{1}{2}+2\sin^{2}\theta_{\mathrm{W}}$ for the gas of electron neutrinos.
If $\nu=\nu_{\mu,\tau}$, one has that $c_{\mathrm{V}}=-\tfrac{1}{2}+2\sin^{2}\theta_{\mathrm{W}}$. Here, $\theta_{\mathrm{W}}$
is the Weinberg angle, with $\sin^{2}\theta_{\mathrm{W}}=0.23$. The
integral in Eq.~(\ref{eq:dotEfin}) is to be evaluated numerically.

\section{Neutrino cluster cooling}\label{sec:CLUSTER}

In this section, we apply the results of Sec.~\ref{sec:CHEREMGEN}
to estimate the cooling rate of a neutrino cluster.

We assume that a neutrino cluster is formed in the early universe
owing to a fluctuation. The temperature of background plasma is
$T$ at that time. Our strategy is to find the range of temperatures $T_\mathrm{min} < T < T_\mathrm{max}$ when the cooling rate is faster than the universe expansion. In this situation, a cluster can survive to the present time universe. It is clear that $T_\mathrm{max} \lesssim T^{(\nu)}_\text{decoupl}\sim(2-3)\,\text{MeV}$, which is the neutrino decoupling temperature~\cite[pp.~22--23]{GorRub11}. In this case, a cluster is not destroyed by thermal fluctuations in the outer neutrino gas.

Our calculations are based on the analytical dispersion relation for a plasmon, which is known either in nonrelativistic or in ultrarelativistic cases. We shall see shortly that $T_\mathrm{min}$ corresponds to the nonrelativistic plasma. That is why, we restrict ourselves to the range $T_\mathrm{min} < T < 500\,\text{keV}$ if we study a cluster composed of $\nu_e$. Since the formation of a cluster is a random process, it desirable to have $T_\mathrm{min}$ as low as possible. In such a situation, clusters have more chances for the appearance. Hence, a greater number of clusters can be created in the whole universe. The interval $500\,\text{keV} < T < T^{(\nu)}_\text{decoupl}$ is not covered by our study since it is beyond the scope of the approximation adopted.

\begin{figure}
  \centering
  \includegraphics[scale=0.4]{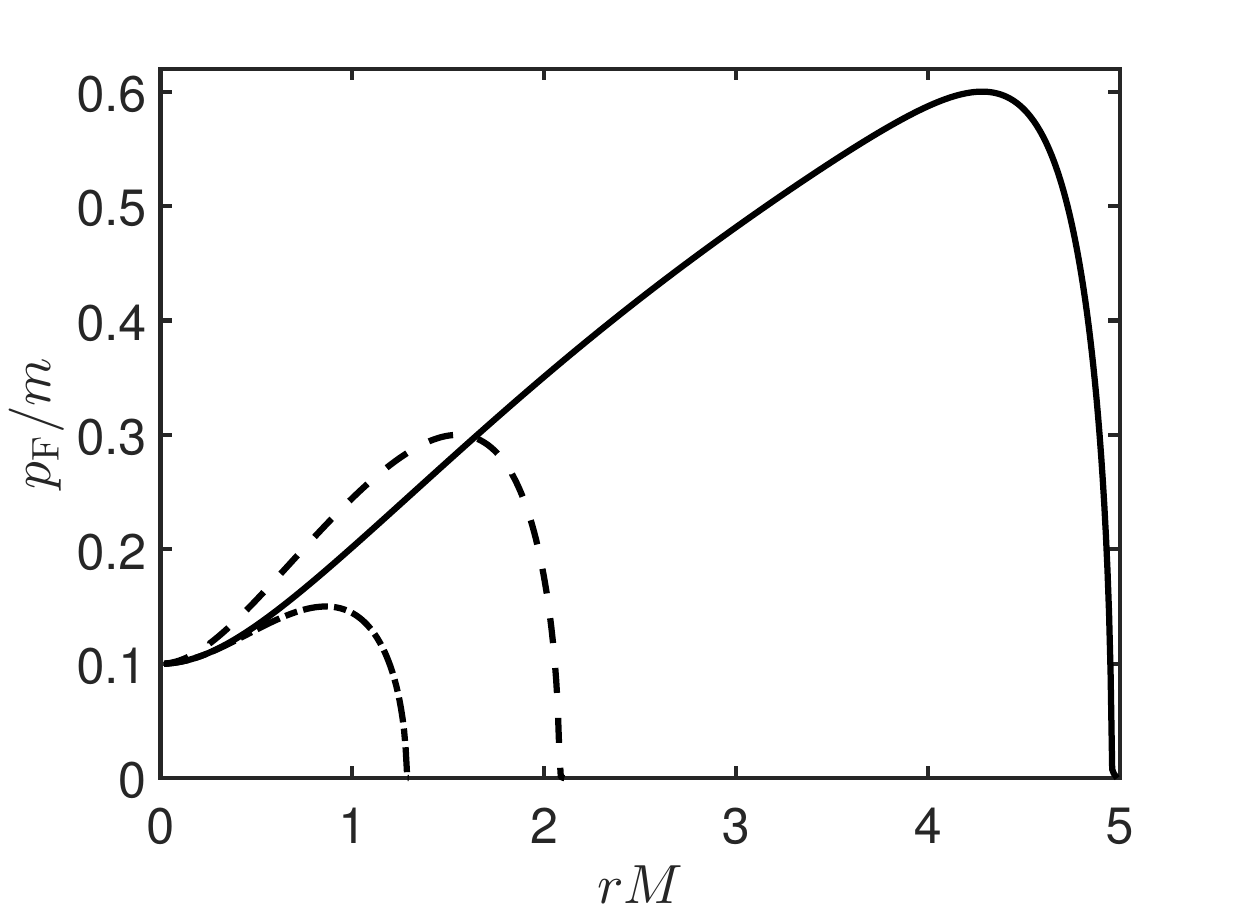}
  \protect
  \caption{The distribution of the Fermi momentum inside the neutrino clusters. In
this plot, $m\equiv m_{\nu}=0.1\,\text{eV}$ is the neutrino mass
and $M\equiv m_{s}$ is the scalar particle mass. The solid line corresponds
to $\mu_{\nu}^{(\mathrm{now})}=0.6m_{\nu}$, $R_{\mathrm{now}}\approx5m_{s}^{-1}$
and $p_{\mathrm{F}}^{(\mathrm{max})}\approx0.6m_{\nu}$; the dashed
line corresponds to $\mu_{\nu}^{(\mathrm{now})}=0.3m_{\nu}$, $R_{\mathrm{now}}\approx2.1m_{s}^{-1}$
and $p_{\mathrm{F}}^{(\mathrm{max})}\approx0.3m_{\nu}$; and dash-dotted
line corresponds to $\mu_{\nu}^{(\mathrm{now})}=0.15m_{\nu}$, $R_{\mathrm{now}}\approx1.3m_{s}^{-1}$
and $p_{\mathrm{F}}^{(\mathrm{max})}\approx0.15m_{\nu}$. All cluster
parameters are given in the present time universe. The figure is taken
from Ref.~\cite{Dvo24}.\label{fig:clust}}
\end{figure}

Since the neutrino gas becomes denser in a cluster,
its temperature is higher than $T$: $T_{\mathrm{clust}}>T$. To find
the relation between $T_{\mathrm{clust}}$ and $T$, we should consider
a particular cluster. The numerical simulations of the cluster structure were carried out
in Ref.~\cite{Dvo24} for different parameters of the neutrino gas
and scalar particles. The samples of these simulations are present
in Fig.~\ref{fig:clust}. We consider in details one of the clusters
depicted in Fig.~\ref{fig:clust} by the solid line. It has the chemical
potential $\mu_{\nu}^{(\mathrm{now})}=0.6m_{\nu}$, the radius $R_{\mathrm{now}}\approx5m_{s}^{-1}$,
and the maximal Fermi momentum $p_{\mathrm{F}}^{(\mathrm{max})}\approx0.6m_{\nu}$.
These cluster parameters correspond to the present time universe.
We take that the neutrino mass is $m_{\nu}=0.1\,\text{eV}$.

Since neutrinos in a cluster are degenerate in the present
universe, the neutrino density inside a cluster now has the value
\begin{equation}
  n_{\mathrm{clust}}^{(\mathrm{now})}
  =2\int_{0}^{p_{\mathrm{F}}^{(\mathrm{max})}}\frac{\mathrm{d}^{3}P}{(2\pi)^{3}}
  =9.5\times10^{8}\,\text{cm}^{-3}.
\end{equation}
Here we consider the upper bound for the neutrino density taking $p_{\mathrm{F}}=p_{\mathrm{F}}^{(\mathrm{max})}=6\times10^{-2}\,\text{eV}$
in the whole cluster. The density of cosmic neutrinos in the present
universe is $n^{(\mathrm{now})}=56\,\text{cm}^{-3}$. We assume that
a cluster evolves in time, with its size being affected only by the
universe expansion. Thus, one has the following relation between $n_{\mathrm{clust}}^{(\mathrm{now})}$
and $n^{(\mathrm{now})}$, as well as $n_{\mathrm{clust}}$ and $n$
which are the densities of neutrinos inside a cluster and of background
neutrinos at the time of the cluster formation:
\begin{equation}\label{eq:nratio}
  \frac{n_{\mathrm{clust}}}{n}=\frac{n_{\mathrm{clust}}^{(\mathrm{now})}}{n^{(\mathrm{now})}}=1.7\times10^{7}.
\end{equation}
To get the upper bound on $T_{\mathrm{clust}}$ we assume that a cluster is formed quite quickly, without the energy transfer to the outer medium. It corresponds to an adiabatic process, with the
heat capacity ratio $\gamma=5/3$. Thus, one gets the relation between $T_{\mathrm{clust}}$
and $T$,
\begin{equation}\label{eq:TclustT}
\frac{T_{\mathrm{clust}}}{T}=\left(\frac{n_{\mathrm{clust}}}{n}\right)^{\gamma-1}=6.6\times10^{4},
\end{equation}
where we use Eq.~(\ref{eq:nratio}).

Since the neutrino distribution functions in Eq.~(\ref{eq:dotEfin})
correspond to $T_{\mathrm{clust}}$, it is convenient to normalize
the integration variables $E_{1}$ and $\omega$ to $T_{\mathrm{clust}}$,
with $x=E_{1}/T_{\mathrm{clust}}$ and $y=\omega/T_{\mathrm{clust}}$
being dimensionless parameters. Finally, we rewrite Eq.~(\ref{eq:dotEfin})
in the form,
\begin{align}\label{eq:dotEdimless}
  \dot{E}= & \frac{R^{3}G_{\mathrm{F}}^{2}c_{\mathrm{V}}^{2}T_{\mathrm{clust}}^{9}}{24\pi^{3}\alpha}I,
  \nonumber
  \\
  I= & \int_{0}^{\infty}\frac{\mathrm{d}x}{e^{x-\xi}+1}
  \int_{y_{1}}^{y_{2}} \frac{ y^{5}\mathrm{d}y }{ e^{- |x-y| + \xi }+1}
  \left(
    1-\frac{y^{2}}{y_{k}^{2}}
  \right)^{3}
  \left[
    2x(x-y)+\frac{1}{2}
    \left(
      y^{2}-y_{k}^{2}
    \right)
  \right],
\end{align}
where $y_{k}=y_{k}(y)$ obeys the relation $y^{4}=b^{2}(y^{2}+ay_{k}^{2})$
(see Eq.~(\ref{eq:disprelLfin})), $y_{1}=b\sqrt{1+a}$, $y_{2}=y_{2}(x)$
is the maximal real root of the equation $y_{k}+y=2x$ (see Eq.~(\ref{eq:omegamaxeq})),
$\xi=\mu_{\nu}/T_{\mathrm{clust}}$, $a=3T/m$, $b=\omega_{p}/T_{\mathrm{clust}}$,
and $\alpha=\tfrac{e^{2}}{4\pi}=7.3\times10^{-3}$ is the fine structure
constant. We remind that we deal with a nonrelativistic primordial
plasma. Thus, we consider the range of temperatures $T\lesssim m$.

The dimensionless parameter $b$ is temperature dependent, $b=\omega_{p}/T_{\mathrm{clust}}\propto\sqrt{T}$.
Using Eq.~(\ref{eq:TclustT}), we write down that
\begin{equation}\label{eq:bfin}
b(T)=2.5\times10^{-12}\left(\frac{T}{\text{keV}}\right)^{1/2}.
\end{equation}
To obtain Eq.~(\ref{eq:bfin}) we take that the electron density
at the cluster formation reads $n_{e}(T)=n_{e}^{(\mathrm{now})}\frac{T^{3}}{T_{\mathrm{CMB}}^{3}}$,
where the present time electron density $n_{e}^{(\mathrm{now})}=\frac{\rho_{\mathrm{baryon}}}{m_{p}}=2.5\times10^{-7}\,\text{cm}^{-3}$,
the present time density of baryonic matter $\rho_{\mathrm{baryon}}\approx4.2\times10^{-31}\,\text{g}\cdot\text{cm}^{-3}$,
which is about $(4-5)\,\%$ of the critical density $\rho_{c}\approx9.4\times10^{-30}\,\text{g}\cdot\text{cm}^{-3}$,
$T_{\mathrm{CMB}}=2.7\,\text{K}$ is the temperature of the cosmic
microwave background radiation, and $m_{p}$ is the proton mass. Here,
we assume that, roughly, the present time universe consists of the
electroneutral hydrogen plasma.

In Eq.~(\ref{eq:dotEdimless}), we assume that plasmons carry away
the energy from the entire cluster. However, Cherenkov plasmons are
unstable. It results from the nonzero imaginary part of $\Pi_{\mathrm{L}}$, which is studied in Appendix~\ref{sec:DISPREL}. Hence, if emitted plasmon decays inside
a cluster, its energy is returned back to medium. We have to estimate
the propagation length $L$ of a plasmon. The total cooling rate of
a cluster depends whether $L$ is greater or smaller than the cluster
radius $R$. We obtained in Ref.~\cite{Dvo25} that $L\ll R$ for the 
ultrarelativistic plasma. In that situation, we supposed that a cluster
cools down layer-by-layer. It significantly increased the cooling
time. Now, we revisit this issue in case of the nonrelativistic plasma.

We estimate $L$ as $L=v_{\mathrm{ph}}t$, where $v_{\mathrm{ph}}=\mathrm{d}\omega/\mathrm{d}K$
is the plasmon phase velocity, $t\approx1/\delta\omega$ is the typical
time of the plasmon propagation, and $\delta\omega$ is the imaginary
part of the frequency given in Eq.~(\ref{eq:domegaL}). Therefore,
the ratio $\rho=R/L$ reads
\begin{equation}\label{eq:rhoRL}
  \rho=\sqrt{\pi}R
  \left(
    \frac{m}{2T}
  \right)^{3/2}
  \frac{\mathrm{d}K}{\mathrm{d}\omega}\frac{\omega_{p}^{2}}{K^{5}}(K^{2}-\omega^{2})^{2}e^{-\frac{m\omega^{2}}{2TK^{2}}},
\end{equation}
where
\begin{equation}
  \frac{\mathrm{d}K}{\mathrm{d}\omega}=\frac{m}{3T}\frac{\omega}{K}
  \left(
    2\frac{\omega^{2}}{\omega_{p}^{2}}-1
  \right),
\end{equation}
is calculated based on Eq.~(\ref{eq:disprelLfin}). Note that, at
$\omega=\omega_{\mathrm{min}}$, $K=\omega$ and $\rho \to 0$. Thus,
$L\gg R$. 

Then, using Eq.~(\ref{eq:disprelLfin}), rewrite $\rho$ in Eq.~(\ref{eq:rhoRL})
in the form,
\begin{equation}\label{eq:rhox}
  \rho=\sqrt{\pi}\left(\frac{3}{2}\right)^{3/2}\frac{R\omega}{a^{3/2}}\frac{(2\chi^{2}-1)(\chi^{2}-1-a)^{2}}{\chi^{2}(\chi^{2}-1)^{3}}
  \exp\left[-\frac{3}{2(\chi^{2}-1)}\right],
\end{equation}
where $\chi=\omega/\omega_{p}$ and $a=3T/m\ll1$. We estimate the
typical plasmon energy as $\omega\sim T_{\mathrm{clust}}$. Thus,
$\chi=T_{\mathrm{clust}}/\omega_{p}=b^{-1}\gg1$, where $b$ is given
in Eq.~(\ref{eq:bfin}). Therefore, we get the limiting expression
for $\rho$ in Eq.~(\ref{eq:rhox}),
\begin{equation}\label{eq:rholim}
  \rho\to2\sqrt{\pi}\left(\frac{3}{2}\right)^{3/2}RT_{\mathrm{clust}}\frac{b^{2}}{a^{3/2}}
  =4.5\times10^{-14}\left(\frac{T}{\text{keV}}\right)^{-1/2},
\end{equation}
where we use Eqs.~(\ref{eq:TclustT}) and~(\ref{eq:bfin}), as well
as take that $R_{\mathrm{now}}=5m_{s}^{-1}$ and $m_{s}=10^{-4}\,\text{eV}$.
Note that the cluster radius at the time of its formation, $R$, is
related to $R_{\mathrm{now}}$ by $R=R_{\mathrm{now}}T_{\mathrm{CMB}}/T$.
Even if $T\sim\text{a few keV}$, one can see that $\rho\ll1$ in
Eq.~(\ref{eq:rholim}). Hence, $R\ll L$ and the approximation implied
in Eq.~(\ref{eq:dotEdimless}) is valid.

To lower the cluster temperature $T_{\mathrm{clust}}$ to the value
$T$ outside, the cooling time should be less than the universe age
$H^{-1}$, where $H=T^{2}/M_{\mathrm{Pl}}^{*}$ is the Hubble parameter,
$M_{\mathrm{Pl}}^{*}=M_{\mathrm{Pl}}/1.66\sqrt{g_{*}}$, $M_{\mathrm{Pl}}=1.2\times10^{19}\,\text{GeV}$
is the Planck mass, and $g_{*}$ is the number of the relativistic
degrees of freedom. Since $\dot{E}$ in Eq.~(\ref{eq:dotEdimless})
is the emissivity of the entire cluster, the cooling time can be estimated
as $t_{\mathrm{cool}}\sim E_{\mathrm{clust}}/\dot{E}$, where $E_{\mathrm{clust}}=N_{\nu}\bar{E}_{\nu}$
is the energy of all neutrinos in the cluster at the moment of its
formation, $N_{\nu}=n_{\mathrm{clust}}V$ is the number of neutrinos
inside the cluster and $\bar{E}_{\nu}\sim T_{\mathrm{clust}}$ is
the mean neutrino energy in the cluster.

Thus, we should demonstrate that the parameter
\begin{equation}\label{eq:Fdef}
  \Xi=\frac{N_{\nu}T_{\mathrm{clust}}H}{\dot{E}}=\frac{1.1\times10^{-22}}{I(T)}\left(\frac{T}{\text{keV}}\right)^{-3},
\end{equation}
is less than one. In Eq.~(\ref{eq:Fdef}), we account for Eq.~(\ref{eq:TclustT}),
take that $g_{*}\approx10$ at $T<1\,\text{MeV}$~\cite[p.~409]{GorRub11}
and $c_{\mathrm{V}}\approx1$ for electron neutrinos. We also take
into account that $n_{\mathrm{clust}}=9.5\times10^{8}\,\text{cm}^{-3}\tfrac{T^{3}}{T_{\mathrm{CMB}}^{3}}$,
which results from Eq.~(\ref{eq:nratio}).

The simulation of the neutrino clusters structure in Ref.~\cite{Dvo24}
implies that the neutrino chemical potential is nonzero inside the
cluster. For example, here, we consider the cluster with $\mu_{\nu}^{(\text{now})}=0.6m_{\nu}=6\times10^{-2}\,\text{eV}$.
Thus, the parameter $\xi=\mu_{\nu}/T_{\text{clust}}$ in Eq.~(\ref{eq:dotEdimless})
is $|\xi|=3.9\times10^{-3}$. We suppose that both positive and negative
neutrino chemical potential are possible since it is unknown whether
the cluster consists of particles or antiparticles. Note that our
value of the neutrino asymmetry parameter is greater than the upper
bound on the global asymmetry, normalized by $T_{\text{clust}}$,
established in Ref.~\cite{Man12}, $|\xi_{\text{our}}|>\tfrac{|\mu_{\nu}^{(\text{global})}|}{T_{\text{clust}}}=\tfrac{|\mu_{\nu}^{(\text{global})}|}{6.6\times10^{4}T}\approx2.2\times10^{-6}$.
The global value corresponds to $\eta=\tfrac{|n_{\nu}-n_{\bar{\nu}}|}{n_{\gamma}}\sim0.1$.
This fact should not be discouraging since the neutrino density is
raised locally only inside a cluster.

\begin{figure}
  \centering
  \includegraphics[scale=0.4]{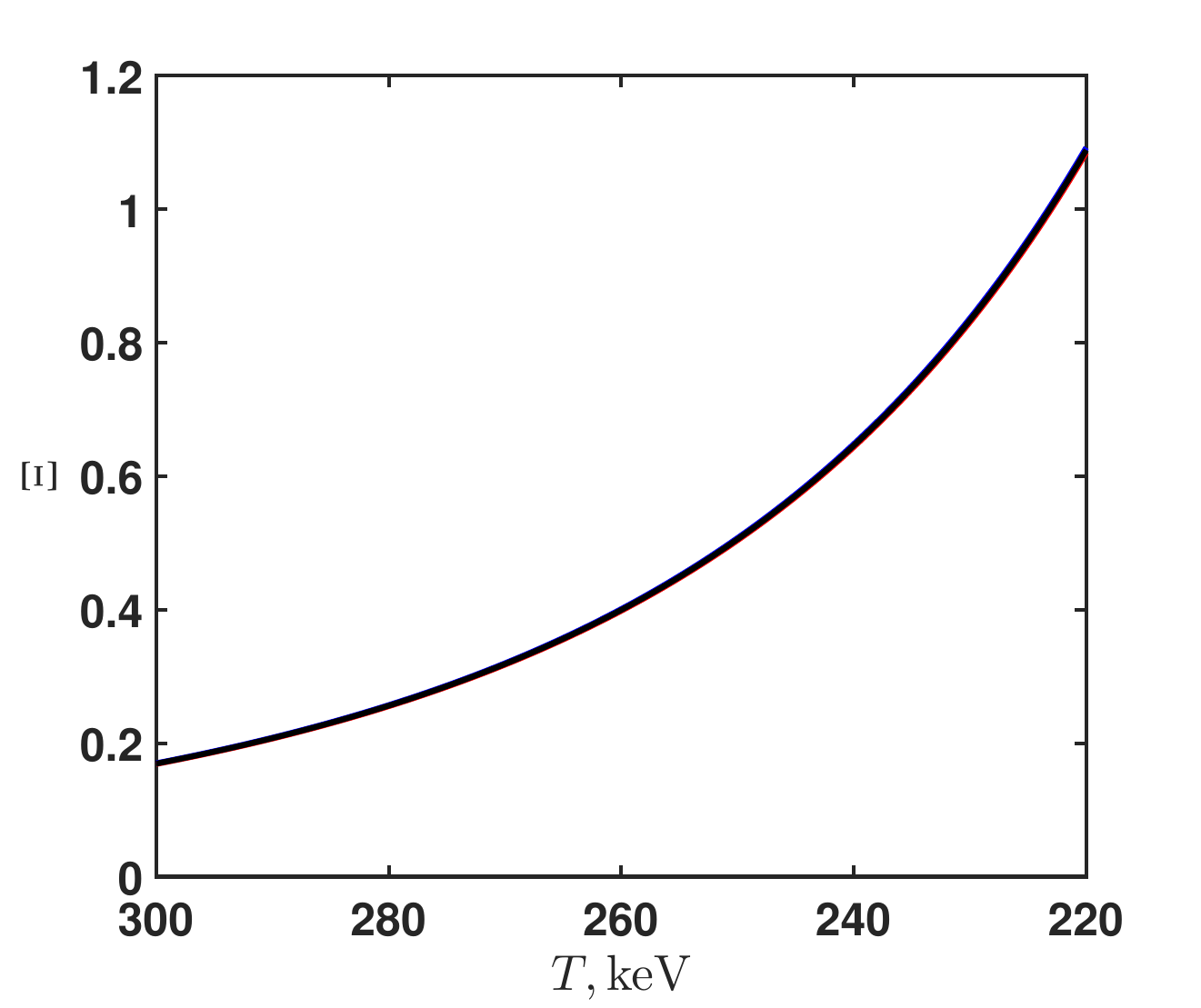}
  \protect
  \caption{The cooling parameter $\Xi$ in Eq.~(\ref{eq:Fdef}) versus $T$
for various asymmetry parameters, $\xi=\pm3.9\times10^{-3}$ (red
and blue lines) and $\xi=0$ (black line). The curves for different
$\xi$ almost overlap. This cooling corresponds to the cluster shown
in Fig.~\ref{fig:clust} by the solid line. It has $\mu_{\nu}^{(\mathrm{now})}=0.6m_{\nu}$,
$R_{\mathrm{now}}\approx5m_{s}^{-1}$ and $p_{\mathrm{F}}^{(\mathrm{max})}\approx0.6m_{\nu}$.\label{fig:coolparam}}
\end{figure}

In Fig.~\ref{fig:coolparam}, we show the evolution of the cooling
parameter $\Xi$ versus $T$ for the cluster shown by the solid line
in Fig.~\ref{fig:clust}. The range of temperatures is $220\,\text{keV}<T<300\,\text{keV}$.
We consider three cases: $\xi=\pm3.9\times10^{-3}$, shown by red
and blue lines, and $\xi=0$, depicted by the black line. One can
see that the cooling parameters are almost indistinguishable in these
situations, i.e. the nonzero chemical potential of neutrinos does
not affect the cooling process.

One can see in Fig.~\ref{fig:coolparam} that $\Xi<1$ for $T\gtrsim220\,\text{keV}$.
Thus, if the neutrino cluster with chosen characteristics is formed
in the epoch when $T\gtrsim220\,\text{keV}$, the emission of Cherenkov
plasmons is effective to cool it down to the temperature $T$ of the
outside plasma. It happens since the cooling rate is faster than the
universe expansion at that time. We notice that, in Fig.~\ref{fig:coolparam},
we have $T/m<1$ rather than $T/m\ll1$, which should be in a nonrelativistic
approximation. Thus, relativistic corrections can, in principle, somehow
affect our estimates.

We also estimated the evolution of the cooling parameter for the clusters
shown by dashed and dash-dotted lines in Fig.~\ref{fig:clust}. For
these clusters, it turns out that $\Xi<1$ for $T\gtrsim500\,\text{keV}$.
It means that these clusters can cool down only in the hot universe
where primordial plasma is (ultra-)relativistic. The approximation
of the nonrelativistic plasma, used in our work, is not applicable
in these situations. That is why we do not show such cooling parameters here.
This behavior of $\Xi$ can be explained by the fact that the emissivity
by Cherenkov plasmons is higher at a greater $T_\text{clust}$. Thus, to get the appropriate cooling rate of a cluster with a smaller radius it should be formed in at earlier epoch when the universe temperature is higher.

\section{Discussion}\label{sec:CONCL}

We have studied the emission of Cherenkov plasmons by primordial neutrinos
which form the gas with a nonzero temperature and a chemical potential.
Despite a neutrino is an electrically neutral particle, one can consider
its induced electric charge in medium owing to the loop effects~\cite{OraSemSmo94}.
It is this charge which is responsible for the neutrino Cherenkov
radiation.

Previously, the Cherenkov radiation of neutrinos was discussed mainly in the
context of an individual particle moving through medium~\cite{OliNiePal96,Sah97}.
In our work, we considered a gas of neutrinos emitting plasmons. Thus,
we had to average the matrix element over the incoming and outgoing
neutrino states having the Fermi-Dirac distribution functions.

Our calculations implied the generalized polarization tensor of a
plasmon and the plasmon form factors in medium. We have presented the basic steps in their
computation in Appendices~\ref{sec:SUMFTFT} and~\ref{sec:CALCPLASMFF}
using the imaginary time perturbation theory. We have decided to provide
these computations, firstly, for the convenience of a reader. Secondly,
as a rule, the plasmon polarization tensor is calculated in the Hard Thermal Loops limit (see, e.g., Ref.~\cite[pp.~118--124]{Bel04}),
which is valid mainly for an ultrarelativistic plasma.
Moreover, in this limit, one does not take into account the chemical potential contribution. In our work,
we consider the situation of nonrelativistic plasma when the density
of charged leptons, or their chemical potential, is important.

In Sec.~\ref{sec:CHEREMGEN}, we have derived the general expression
for the neutrino gas emissivity. It was found that only longitudinal
plasmons contribute to the matrix element. Previously, this fact was
mentioned in Ref.~\cite{Sah97} in case of the ultraretivistic plasma.
We have also established the range for the integration over the plasmon
frequencies.

Then, in Sec.~\ref{sec:CLUSTER}, we have considered the application
of our results for the cooling of a neutrino cluster formed in the early
universe. This kind of neutrino clusters was described, first, in
Ref.~\cite{SmiXu22}. Neutrinos are held together inside a cluster
by the exchange of hypothetical light scalar particles. However, when
a cluster is formed, the neutrino gas is compressed and heats up.
Hence, a cluster can be destroyed by thermal effects. The cooling
mechanisms put forward in Ref.~\cite{SmiXu22} turned out to be inefficient.

In Ref.~\cite{Dvo24}, we proposed the mechanism for the cluster
cooling based on the emission of Cherenkov plasmons by neutrinos in
a cluster. Nevertheless, the consideration of this process was quite estimatory
in Ref.~\cite{Dvo24}. Indeed, to evaluate the neutrino gas emissivity in Ref.~\cite{Dvo24}, we used the emission rate of plasmons derived in Ref.~\cite{OliNiePal96}. It was found in Ref.~\cite{OliNiePal96} that the contribution of transverse plasmons to the emission rate dominates over that of longitudinal ones. In the present work, we have obtained that the condition for the Cherenkov plasmon emission, $K>\omega$, can be fulfilled only for longitudinal plasmons, as we mentioned above. Therefore, these results of Ref.~\cite{Dvo24} can be considered as correct only to get an estimate of the order of magnitude.

The description of the cluster cooling by
the mechanism in question was refined in Ref.~\cite{Dvo25}. We used
the model of the ultrerelativistic plasma Ref.~\cite{Dvo25}. However,
the temperature interval where the cooling is efficient, obtained
in Ref.~\cite{Dvo25}, spanned both relativistic and nonrelativistic
ranges. The extension of the temperature range for the nonrelativistic plasma was not justified in Ref.~\cite{Dvo25}. Moreover, the condition of the longitudinal Cherenkov plasmons emission, $K>\omega$, is satisfied formally at $K > K_1 = \infty$ for the relativistic plasma (see, e.g., Refs.~\cite{BraSeg93} and~\cite[pp.~215--216]{Raf96}). Thus, the phase space for the plasmon emission in Eq.~\eqref{eq:dotEdef} is quite small in this situation. It leads to the reduced emissivity of the neutrino gas compared to the nonrelativistic plasma. We also mention that the concept of the layer-by-layer cluster cooling, used in Ref.~\cite{Dvo25}, is unlikely to be valid since plasmon propagation length was underestimated in Ref.~\cite{Dvo25}.

We also make a general comment on the previous applications of the Cherenkov emission by neutrinos. This effect was used, e.g., in Refs.~\cite{OliNiePal96,Sah97} to estimate the energy losses by a beam of ultrarelativistic neutrinos, originated, for instance, in a SN explosion. Thus, the momentum of an incoming neutrino was fixed. Moreover, refraction index $n_\gamma=K/\omega$ was taken to have a fixed value. These assumptions are justified if one needs to obtain an estimate for the neutrino emissivity. In the present work, we deal with the neutrino gas characterized by the temperature of neutrinos $T_\nu$ and their chemical potential $\mu_\nu$. Therefore, we have to express $\dot{E}$ in Eq.~\eqref{eq:dotEdef} through $T_\nu$ and $\mu_\nu$ by averaging over the incoming and outgoing neutrino states. Our expression for $\dot{E}$ absorbs the dependence on $n_\gamma$ since we numerically integrate over $\omega$ in Eq.~\eqref{eq:dotEdimless} accounting for the dispersion relation $K(\omega)$ for longitudinal plasmons.

In the present work, we have reconsidered the cluster cooling in case
when plasma is nonrelativistic, $T<m$. This situation is of the main
importance since neutrinos are sure to decouple from the primordial
plasma at that epoch. It happens
at $\sim(2-3)\,\text{MeV}$. Therefore
a cluster is not destroyed by thermal effects associated with the neutrino-plasma interactions.

Using particular cluster parameters, obtained in Ref.~\cite{Dvo24},
we have found that the Cherenkov plasmon emission is, indeed, efficient
to cool down some clusters since the cooling rate is higher than the
universe expansion. It happens for clusters with greater radii and higher internal densities. We
have found that the proposed cooling mechanism is valid if a cluster
is formed at $T\gtrsim220\,\text{keV}$. This range covers the neutrino decoupling temperature. We have also found that the nonzero chemical
potential of neutrinos does not contribute significantly to the cluster
cooling rate. 

In our work, we studied the case of electron neutrinos interacting
with a hydrogen plasma. In principle, a cluster can consist of other
flavor neutrinos. However, the proposed cooling mechanism is not efficient
for $\nu_{\mu}$ and $\nu_{\tau}$. Indeed, the vector coupling constant,
which $\dot{E}$ in Eq.~(\ref{eq:dotEfin}) is proportional to, is
much smaller for other flavor neutrinos, $c_{\mathrm{V}}^{(\nu_{\mu},\nu_{\tau})}\ll c_{\mathrm{V}}^{(\nu_{e})}$.

In summary, the main problem, solved in the present work, was the determination of the lower bound for the temperature $T_\text{crit} \approx 220\,\text{keV}$ of the early universe when a neutrino cluster can be formed. We have obtained that, if a cluster is formed in the temperature range $T_\text{crit} < T < T^{(\nu)}_\text{decoupl}$, it can potentially survive to the present time universe. Since the cluster formation is a random process affected by many factors, our main goal is to obtain the interval $(T_\text{crit}, T^{(\nu)}_\text{decoupl})$ as wide as possible to increase chances for a cluster to appear. The previous estimates for $T_\text{crit}$ in Refs.~\cite{Dvo24,Dvo25} are unlikely to be reliable.

\appendix

\section{Sum rules for the thermal loop integrals}\label{sec:SUMFTFT}

In this Appendix, we provide the computation of
the sums over the Matsubara frequencies for some integrals one deals
with in the calculation of the generalized polarization tensor in
Eq.~(\ref{eq:Pimunudef}). Here, we account for both the temperature
and the chemical potential.

\begin{figure}
  \centering
  \includegraphics[scale=0.2]{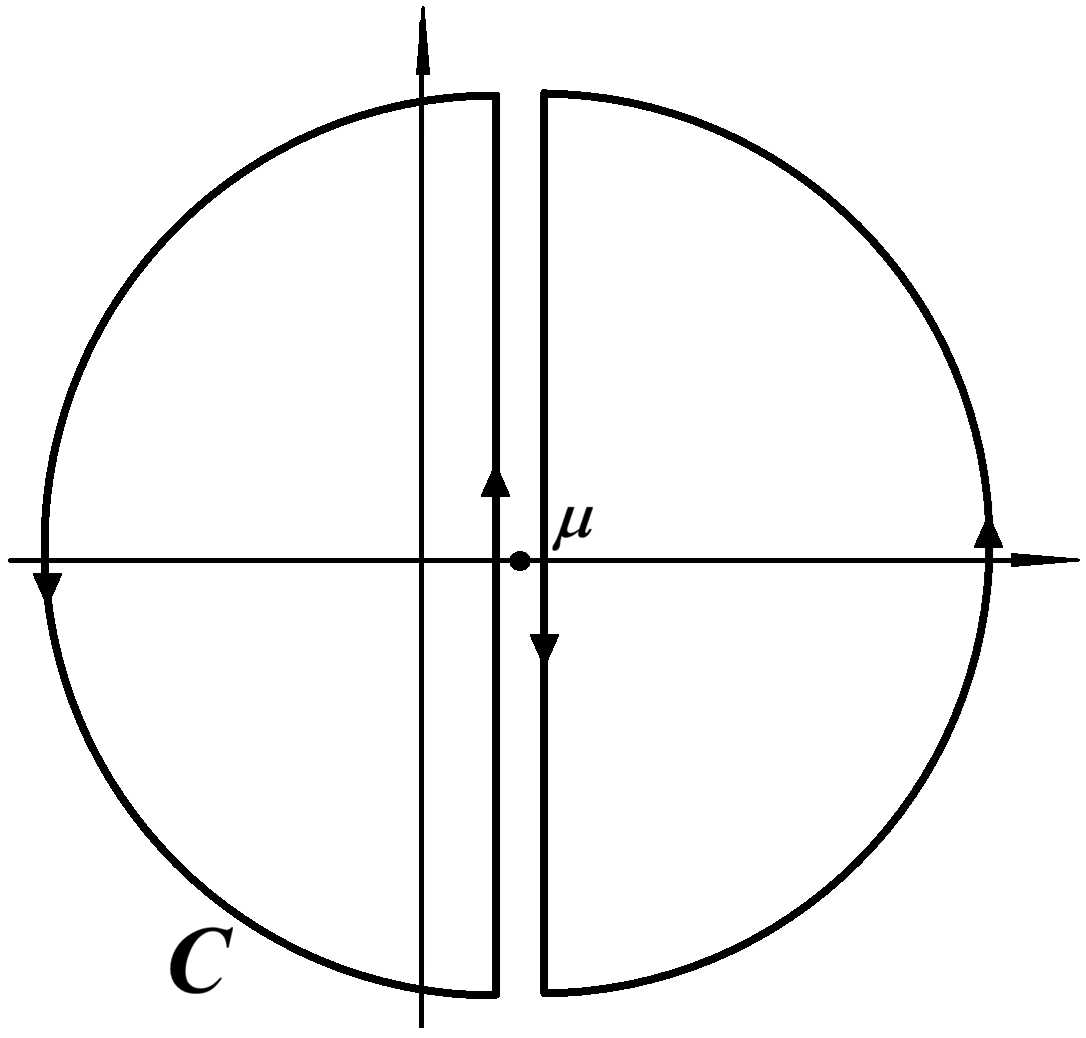}
  \protect
  \caption{The contour $C$ for the integration over the complex variable $P_{0}$
  in Eq.~(\ref{eq:sumMatsfer}).\label{fig:cont}}
\end{figure}

The contribution of the nonzero temperature $T$ and the chemical
potential $\mu$ to the 4D loop integration is accounted for by the replacement,
\begin{equation}\label{eq:Tmurule}
  \int\frac{\mathrm{d}^{4}P}{(2\pi)^{4}}f(P_{0},\mathbf{P})\to\int\frac{\mathrm{d}^{3}P}{(2\pi)^{3}}T\sum_{n}
  f(P_{0}=\mathrm{i}\omega_{n}+\mu,\mathbf{P}),
\end{equation}
for any function $f(P^\mu)$. The sum over the fermionic Matsubara frequencies
$\omega_{n}=2\pi(n+1)/\beta$ in Eq.~(\ref{eq:Tmurule}) is calculated
as
\begin{equation}\label{eq:sumMatsfer}
  T\sum_{n=-\infty}^{+\infty}f(P_{0}=\mathrm{i}\omega_{n}+\mu)
  =-\frac{1}{\beta}\int_{C}\frac{\mathrm{d}P_{0}}{2\pi\mathrm{i}}f(P_{0})\frac{\beta}{2}\tanh(\beta[P_{0}-\mu]/2),
\end{equation}
where $\beta=1/T$ is the reciprocal temperature and the contour $C$
is shown in Fig.~\ref{fig:cont}.

Equations~\eqref{eq:Tmurule} and~\eqref{eq:sumMatsfer} are applied for the computation of the longitudinal form factor $\Pi_\mathrm{L}$, which is defined shortly in Appendix~\ref{sec:PLASMPROP}. In particular, one has the following loop integrals in the generalized photon
polarization operator in Eq.~(\ref{eq:Pimunudef}):
\begin{align}\label{eq:S0-S2}
  S_{0} & =\int\frac{\mathrm{d}^{4}P}{(2\pi)^{4}}\frac{1}{
  \left[
    P^{2}-m^{2}
  \right]
  \left[
    (P-k)^{2}-m^{2}
  \right]},
  \nonumber
  \\
  S_{2} & =\int\frac{\mathrm{d}^{4}P}{(2\pi)^{4}}\frac{P_{0}(P_{0}-k_{0})}{
  \left[
    P^{2}-m^{2}
  \right]
  \left[
    (P-k)^{2}-m^{2}
  \right]},
\end{align}
where $m$ is the mass of the fermion in the loop and $k^\mu=(k^{0},\mathbf{K})$
is the external momentum. Using Eqs.~(\ref{eq:Tmurule}) and~(\ref{eq:sumMatsfer}),
one gets the explicit form of the integrals in Eq.~(\ref{eq:S0-S2}),
\begin{align}\label{eq:S0-S2expl}
  S_{0}= & \frac{1}{4}\int\frac{\mathrm{d}^{3}P}{(2\pi)^{3}}
  \bigg[
    \frac{1}{E_{1}}
    \left(
      \frac{2n_{1}-1}{(k_{0}-E_{1}-E_{2})(k_{0}-E_{1}+E_{2})}+\frac{2\bar{n}_{1}-1}{(k_{0}+E_{1}+E_{2})(k_{0}+E_{1}-E_{2})}
    \right)
    \nonumber
    \\
    & +
    \frac{1}{E_{2}}
    \left(
      \frac{2n_{2}-1}{(k_{0}+E_{1}+E_{2})(k_{0}-E_{1}+E_{2})}+\frac{2\bar{n}_{2}-1}{(k_{0}-E_{1}-E_{2})(k_{0}+E_{1}-E_{2})}
    \right)
  \bigg],
  \nonumber
  \\
  S_{2}= & \frac{1}{4}\int\frac{\mathrm{d}^{3}P}{(2\pi)^{3}}
  \bigg[
    \frac{(E_{1}-k_{0})(2n_{1}-1)}{(k_{0}-E_{1}-E_{2})(k_{0}-E_{1}+E_{2})}+\frac{(E_{1}+k_{0})(2\bar{n}_{1}-1)}{(k_{0}+E_{1}+E_{2})(k_{0}+E_{1}-E_{2})}
    \nonumber
    \\
    & +
    \frac{(E_{2}+k_{0})(2n_{2}-1)}{(k_{0}+E_{1}+E_{2})(k_{0}-E_{1}+E_{2})}+\frac{(E_{2}-k_{0})(2\bar{n}_{2}-1)}{(k_{0}-E_{1}-E_{2})(k_{0}+E_{1}-E_{2})}
  \bigg],
\end{align}
where $E_{1}=\sqrt{\mathbf{P}^{2}+m^{2}}$, $E_{2}=\sqrt{(\mathbf{P}-\mathbf{K})^{2}+m^{2}}$,
$n_{1,2}=\left\{ \exp(\beta[E_{1,2}-\mu])+1\right\} ^{-1}$, and $\bar{n}_{1,2}=\left\{ \exp(\beta[E_{1,2}+\mu])+1\right\} ^{-1}$.
To derive Eq.~(\ref{eq:S0-S2expl}) we take into account that $k^\mu$
corresponds to a boson. Thus, $k_{0}=2\pi\mathrm{i}n/\beta$,
with $n=0,\pm1,\dotsc$.

\section{Plasmon properties in matter}\label{sec:PLASMPROP}

In this Appendix, we list the properties of a plasmon in background
matter. Here, we rely on Ref.~\cite{NiePal89}.

In a nonchiral medium, the plasmon polarization tensor is decomposed in the basis of the projection operators in Eq.~\eqref{eq:Pimunudecom}. The projection
operators in Eq.~(\ref{eq:Pimunudecom}) have the form,
\begin{equation}\label{eq:projoperdef}
  Q_{\mu\nu} =-\frac{k^{2}}{K^{2}}
  \left(
    v_{\mu}-\omega\frac{k_{\mu}}{k^{2}}\right)\left(v_{\nu}-\omega\frac{k_{\nu}}{k^{2}}
  \right),
  \quad
  R_{\mu\nu} = g_{\mu\nu}-\frac{k_{\mu}k_{\nu}}{k^{2}}-Q_{\mu\nu},
\end{equation}
where we choose the frame where the background matter is at rest
having the four velocity $v^{\mu}=(1,\mathbf{0})$. It is instructive to write down the components of the projection operators
in Eq.~(\ref{eq:projoperdef}),
\begin{align}\label{eq:projoperprop}
  Q_{00} & =-\frac{K^{2}}{k^{2}},
  \quad
  Q_{0i}=-\frac{\omega k_{i}}{k^{2}},
  \quad
  Q_{ij}=-\frac{\omega^{2}}{k^{2}}\frac{K_{i}K_{j}}{K^{2}},
  \nonumber
  \\
  R_{00} & =0,
  \quad
  R_{0i}=0,
  \quad
  R_{ij}=-\delta_{ij}+\frac{K_{i}K_{j}}{K^{2}}.
\end{align}
Moreover, the projection operators obey the properties,
\begin{equation}\label{eq:projoperorth}
  Q^{2} =Q,
  \quad
  R^{2}=R,
  \quad
  QR=0,
\end{equation}
which are represented in the symbolic form.

The dispersion relations for plasmons are obtained in Ref.~\cite{NiePal89} by considering the poles of the dressed plasmon propagator. One gets that $k^{2}=\Pi_{\mathrm{L,R}}$ for longitudinal and transverse plasmons. If one defines the polarization vectors as $e_{\mathrm{L}}^{\mu}$ and $e_{1,2}^{\mu}$ for longitudinal and transverse plasmons, one has the following rules to sum over the polarizations:
\begin{equation}\label{eq:eLsum}
  e_{\mathrm{L}\mu}e_{\mathrm{L}\nu} = Q_{\mu\nu},
  \quad
  \sum_{s=1,2}e_{s,\mu}^{*}e_{s,\nu}=-R_{\mu\nu},
\end{equation}
which are used, e.g., in Eq.~(\ref{eq:M2sumpol}). We also provide the relations between the plasmon form factors and
the components of $\Pi_{\mu\nu}$,
\begin{equation}\label{eq:PiLPi00}
  \Pi_{\mathrm{L}} =
  \left(
    1-\frac{\omega^{2}}{K^{2}}
  \right)
  \Pi_{00},
  \quad
  \Pi_{\mathrm{T}} =\frac{1}{2}
  \left(
    \Pi_{\mu}^{\mu}-\Pi_{\mathrm{L}}
  \right).
\end{equation}
where we use Eqs.~(\ref{eq:Pimunudecom}), \eqref{eq:projoperdef}, and~(\ref{eq:projoperprop}).

Finally, we mention that the elementary electric charge $e$, which
is the coupling constant between electromagnetic and electron-positron
fields, should be renormalized in background matter. Thus, we should
replace
\begin{equation}\label{eq:renorm}
  e\to e\sqrt{Z},
  \quad
  Z=\frac{\omega}{K}\frac{\mathrm{d}\omega}{\mathrm{d}K}.
\end{equation}
As a rule, the matrix element of a process, involving a plasmon, $\mathcal{M}\propto e$.
Moreover, one typically integrates $|\mathcal{M}|^{2}$ over the plasmon
phase volume $\mathrm{d}^{3}K=K^{2}\mathrm{d}K\mathrm{d}\Omega$;
cf. Eq.~(\ref{eq:dotEdef}). Therefore, the renormalization procedure
in Eq.~(\ref{eq:renorm}) is equivalent to the change of the integration
variable, $Z\mathrm{d}K=\omega\mathrm{d}\omega/K$.

\section{Calculation of the plasmon form factors}\label{sec:CALCPLASMFF}

To calculate the plasmon form factors we can use Eq.~(\ref{eq:Pimunudef}) and put there $c_{\mathrm{V}}=e^{2}$ and $c_{\mathrm{A}} = 0$. In this case, $\tilde{\Pi}_{\mu\nu} \to \Pi_{\mu\nu}$.

We provide some details in the calculation of $\Pi^{00}$. Based on Eq.~(\ref{eq:Pimunudef}),
we obtain that
\begin{align}\label{eq:Pi00gen}
  \Pi^{00}(k_0,\mathbf{K})= & -4 e^2 \int\frac{\mathrm{d}^{4}P}{(2\pi)^{4}}
  \frac{P_{0}(P_{0}-k_{0})+m^{2}+\mathbf{P}(\mathbf{P}-\mathbf{K})}{[P^{2}-m^{2}][(P-k)^{2}-m^{2}]}
  \notag
  \\
  & =
  -4 e^2
  \left(
    S_{2}+S_{0}[\mathbf{P}(\mathbf{P}-\mathbf{K})+m^{2}]
  \right),
\end{align}
where $S_{0}$ and $S_{2}$ are given in Eq.~(\ref{eq:S0-S2expl}).
In Eq.~(\ref{eq:Pi00gen}), the functional notation $S_{0}[\dots]$
means that we include an extra factor under the momentum integration.

We are interested in the matter contribution to the polarization tensor.
That is why, we should subtract the vacuum contribution, which depends
on neither $T$ nor $\mu$, from $\Pi^{\mu\nu}$, $\Pi^{\mu\nu}\to\Pi^{\mu\nu}-\Pi_{\mathrm{vac}}^{\mu\nu}$.
Using Eq.~(\ref{eq:Pi00gen}), we express $\Pi^{00}$ in the form,
\begin{align}\label{eq:Pi00semifin}
  \Pi^{00} = & \frac{e^2}{\pi^{2}}\mathrm{Re}\int_{0}^{\infty}\frac{P^{2}}{E}\mathrm{d}P(n+\bar{n})
  \notag
  \\
  & \times
  \left[
    1+\frac{4Ek_{0}-4E^{2}-k_{0}^{2}+\mathbf{K}^{2}}{4p|\mathbf{K}|}
    \ln
    \left(
      \frac{k_{0}^{2}-\mathbf{K}^{2}-2k_{0}E+2P|\mathbf{K}|}{k_{0}^{2}-\mathbf{K}^{2}-2k_{0}E-2P|\mathbf{K}|}
    \right)
  \right],
\end{align}
where $\mathrm{Re}[f]=[f(k_{0})+f(-k_{0})]/2$. We also omit the index 1 in $n$ and $E$. 

Equation~(\ref{eq:Pi00semifin}) coincides with the result in Ref.~\cite[p.~74]{KapGal06}.
Then, following Ref.~\cite[p.~210]{Raf96},
we consider the approximation where $|k^\mu|\ll P$, with
the fermion mass being kept. To regularize the notations, we take
that a plasmon in on mass shell. Thus, $k^\mu=(\omega,\mathbf{K})$. Using this approximation and Eq.~(\ref{eq:PiLPi00}), we obtain the
longitudinal form factor of a plasmon as
\begin{equation}\label{eq:PiLint}
  \Pi_{\mathrm{L}}=\frac{e^2}{\pi^{2}}
  \left(
    1-\frac{\omega^{2}}{K^{2}}
  \right)
  \int_{0}^{\infty}\frac{\mathbf{P}^2}{E}\mathrm{d}P(n+\bar{n})
  \left[
    1+\frac{\omega^{2}-K^{2}}{\omega^{2}-K^{2}v^{2}}+\frac{\omega}{Kv}\ln
    \left(
      \frac{\omega-Kv}{\omega+Kv}
    \right)
  \right],
\end{equation}
where $v=|\mathbf{P}|/E=|\mathbf{P}|/\sqrt{m^2 + \mathbf{P}^2}$. Equation~\eqref{eq:PiLint} coincides with the result in Ref.~\cite[p.~211]{Raf96} obtained
in frames of the real time perturbation theory.

Finally, we provide the expression for the transverse form factor of a plasmon. Using the same approximations which lead to Eq.~\eqref{eq:PiLint}, one gets that
\begin{equation}\label{eq:PiTint}
  \Pi_{\mathrm{T}}=\frac{e^2}{\pi^{2}}
  \left(
    1-\frac{\omega^{2}}{K^{2}}
  \right)
  \int_{0}^{\infty}\frac{\mathbf{P}^2}{E}\mathrm{d}P(n+\bar{n})
  \left[
    \frac{\omega^{2}}{K^{2}-\omega^{2}}-\frac{\omega}{2Kv}
    \ln
    \left(
      \frac{\omega-Kv}{\omega+Kv}
    \right)
  \right].
\end{equation}
Equation~\eqref{eq:PiTint} is in agreement with the result in Ref.~\cite[p.~211]{Raf96}.

\section{Dispersion relations for plasmons}\label{sec:DISPREL}

In this Appendix, using the results of Appendix~\ref{sec:CALCPLASMFF},
we derive the dispersion relations for plasmons and evaluate the plasmon 
damping.

First, we analyze longitudinal plasmons. We showed in Appendix~\ref{sec:PLASMPROP}
that the dispersion relation results from $\omega^{2}-K^{2}=\Pi_{\mathrm{L}}$.
Using Eq.~(\ref{eq:PiLint}), one gets that the dispersion relation
reads
\begin{equation}\label{eq:disprelLint}
  K^{2}=-\frac{4\alpha}{\pi}\int_{0}^{\infty}\frac{P^{2}}{E}\mathrm{d}P(n+\bar{n})
  \left[
    1+\frac{\omega^{2}-K^{2}}{\omega^{2}-K^{2}v^{2}}+\frac{\omega}{Kv}\ln\frac{\omega-Kv}{\omega+Kv}
  \right],
\end{equation}
where $\alpha=e^{2}/4\pi=7.3\times10^{-3}$ is the fine structure
constant.

To analyze the integral expression in Eq.~(\ref{eq:disprelLint})
we restrict ourselves to the nonrelativistic plasma. The properly
normalized electron distribution function has the form,
\begin{equation}\label{eq:distrnonrel}
  n(P)=\frac{n_{e}}{2}\left(\frac{2\pi}{mT}\right)^{3/2}e^{-\frac{P^{2}}{2mT}},
\end{equation}
where $n_{e}$ is the mean electron density. Moreover, we put $\bar{n}=0$
since the positron distribution is exponentially suppressed, as well as
we take that $P=mv$ and $E=m$. Considering small $v$ in Eq.~\eqref{eq:disprelLint} and using Eq.~\eqref{eq:distrnonrel}, one gets the dispersion relation in question
\begin{equation}\label{eq:disprelLfin}
  \omega^{2}=\omega_{p}^{2}
  \left(
    1+\frac{3T}{m}\frac{K^{2}}{\omega^{2}}
  \right),
\end{equation}
where $\omega_{p}=\sqrt{4\pi\alpha n_{e}/m}$ is the plasma frequency.
In deriving Eq.~(\ref{eq:disprelLfin}), we keep only the leading
terms in the small parameter $T/m$. Note that Eq.~(\ref{eq:disprelLfin})
coincides with the known result, e.g., in Refs.~\cite[p.~213]{Raf96}
and~\cite[p.~134]{PitLif81}.

Besides the real part of $\Pi_\mathrm{L}$ in Eq.~\eqref{eq:PiLint}, this form factor also has an imaginary part. The nonzero $\mathrm{Im}(\Pi_{\mathrm{L}})$ leads to appearance of a small imaginary part of the plasmon frequency $\omega\to\omega+\mathrm{i}\delta\omega$. This the phenomenon is called the Landau damping \cite[pp.~124--127]{PitLif81}. We can evaluate $\delta\omega$ as
\begin{equation}\label{eq:domegaL}
  \delta\omega=\frac{\mathrm{Im}(\Pi_{\mathrm{L}})}{2\omega}
  =\sqrt{\pi}\left(\frac{m}{2T}\right)^{3/2}\frac{\omega_{p}^{2}}{K^{5}}(K^{2}-\omega^{2})^{2}e^{-\frac{m\omega^{2}}{2TK^{2}}}.
\end{equation}
It should be noted that $\delta\omega>0$ in Eq.~(\ref{eq:domegaL})
as predicted in Ref.~\cite[p.~122]{PitLif81}.

Now, we provide the basic results for the transverse plasmons dispersion
relation. Analogously to the longitudinal plasmons case, using Eq.~\eqref{eq:PiTint}, we get that
\begin{equation}\label{eq:disprelT}
  \omega^{2}=K^{2}+\omega_{p}^{2}
  \left(
    1+\frac{T}{m}\frac{K^{2}}{\omega^{2}}
  \right),
\end{equation}
which coincides with the result of Ref.~\cite[p.~213]{Raf96}. The
Landau damping was shown in Ref.~\cite[pp.~130--131]{PitLif81} to
be suppressed for transverse plasmons.


\begin{thebibliography}{50}

\bibitem{Wei20}
  S.~Weinberg,
  Cosmology
  (Oxford University Press, Oxford, 2020).

\bibitem{RosSesTro18}
  L.~Roszkowski, E.~M.~Sessolo, and S.~Trojanowski,
  WIMP dark matter candidates and searches -- current status and future prospects,
  Rep. Prog. Phys. \textbf{81}, 066201 (2018) [arXiv:1707.06277].

\bibitem{Kim87}
  J.~E.~Kim,
  Light pseudoscalars, particle physics and cosmology,
  Phys. Rep. \textbf{150}, 1--177 (1987).

\bibitem{Ber13}
  R.~Bernabei, et al.,
  Final model independent result of DAMA/LIBRA--phase1,
  Eur. Phys. J. C \textbf{73}, 2648 (2013)
  [arXiv:1308.5109].

\bibitem{Boz25}
  N.~Bozorgnia, J.~Bramante, J.~M.~Cline, D.~Curtin, D.~McKeen, D.~E.~Morrissey, A.~Ritz, S.~Viel, A.~C.~Vincent, and Y.~Zhang,
  Dark Matter Candidates and Searches,
  Canadian J. Phys. \textbf{103}, 671--703 (2025)
  [arXiv:2410.23454].

\bibitem{Gun78}
  J.~E.~Gunn, B.~W.~Lee, I.~Lerche, D.~N.~Schramm, and G.~Steigman,
  Some astrophysical consequences of the existence of a heavy stable neutral lepton,
  Astrophys. J. \textbf{223}, 1015--1031 (1978).

\bibitem{Cli06}
  J.~M.~Cline, G.~Herrera, and J.-S.~Roux,
  Neutrinos as Dark Matter
  [arXiv:2603.28859].

\bibitem{Ste98}
  G.~J.~Stephenson, Jr., J.~T.~Goldman, and B.~H.~J.~McKellar,
  Neutrino clouds,
  Int. J. Mod. Phys. A \textbf{13}, 2765--2790 (1998)
  [hep-ph/9603392].

\bibitem{SmiXu22}
  A.~Yu.~Smirnov and X.-J.~Xu,
  Neutrino bound states and bound systems,
  J. High Energy Phys. \textbf{08} (2022) 170
  [arXiv:2201.00939].

\bibitem{Kap04}
  J.~I.~Kapusta,
  Neutrino Superfluidity,
  Phys. Rev. Lett. \textbf{93}, 251801 (2004)
  [hep-th/0407164].

\bibitem{Aza11}
  M.~Azam, J.~R.~Bhatt, and U.~Sarkar,
  Experimental signatures of cosmological neutrino condensation,
  Phys. Lett. B \textbf{697}, 7--10 (2011)
  [arXiv:1008.5214].

\bibitem{Dvo24}
  M.~Dvornikov,
  Superfluidity in neutrino clusters,
  J. Phys. G: Nucl. Part. Phys. \textbf{51}, 075201 (2024)
  [arXiv:2310.04806].

\bibitem{Add22}
  A.~Addazi, S.~Capozziello, Q.~Gan, and A.~Marcian\`{o},
  Dark energy and neutrino superfluids,
  Phys. Dark Univ. \textbf{37}, 101102 (2022)
  [arXiv:2208.03591].

\bibitem{Cap26}
  A.~Capolupo, I.~De~Martino, S.~Monda, R.~Della Monica, and A.~Quaranta,
  Formation and relaxation of halos in the context of wave DM particles evolving on a background of neutrino condensate,
  Phys. Rev. D \textbf{113}, 083040 (2026)
  [arXiv:2603.15719].

\bibitem{Dvo25}
  M.~S.~Dvornikov,
  Neutrino Cluster Cooling by Cherenkov Plasmon Emission,
  Phys. Part. Nucl. \textbf{56}, 472--477 (2025).

\bibitem{OraSemSmo94}
  V.~N.~Oraevsky, V.~B.~Semikoz, and Ya.~A.~Smorodinsky,
  Electrodynamics of neutrino in a medium,
  Phys. Part. Nucl. \textbf{25}, 312--376 (1994).

\bibitem{OliNiePal96}
  J.~C.~D'Olivo, J.~F.~Nieves, and P.~B.~Pal,
  Cherenkov radiation by massless neutrinos,
  Phys. Lett. B \textbf{365}, 178--184 (1996).

\bibitem{Sah97}
  S.~Sahu,
  Cherenkov radiation of longitudinal photons by neutrinos,
  Phys. Rev. D \textbf{56}, 1688--1691 (1997)
  [hep-ph/9612375].

\bibitem{NiePal89}
  J.~F.~Nieves and P.~B.~Pal,
  $P$- and $CP$-odd terms in the photon self-energy within a medium,
  Phys. Rev. D \textbf{39}, 652--659 (1989).

\bibitem{BerLifPit82}
  V.~B.~Berestetskii, E.~M.~Lifshitz, and L.~P.~Pitaevskii,
  Quantum Electrodynamics
  (Pergamon Press, Oxford, 1982), 2nd ed.

\bibitem{GorRub11}
  D.~S.~Gorbunov and V.~A.~Rubakov,
  Introduction to the Theory of the Early Universe: Hot Big Bang Theory
  (World Scientific, Singapore, 2011).

\bibitem{Man12}
  G.~Mangano, G.~Miele, S.~Pastor, O.~Pisanti, and S.~Sarikas,
  Updated BBN bounds on the cosmological lepton asymmetry for non-zero $\theta_{13}$,
  Phys. Lett. B \textbf{708}, 1--5 (2012)
  [arXiv:1110.4335].
  
\bibitem{Bel04}
  M.~Le~Bellac,
  Thermal Field Theory
  (Cambridge University Press, Cambridge, 2004).

\bibitem{BraSeg93}
  E.~Braaten and D.~Segel,
  Neutrino energy loss from the plasma process at all temperatures and densities,
  Phys. Rev. D \textbf{48}, 1478--1491 (1993)
  [hep-ph/9302213].

\bibitem{Raf96}
  G.~G.~Raffelt,
  Stars as Laboratories for Fundamental Physics:
  The Astrophysics of Neutrinos, Axions, and Other Weakly Interacting Particles
  (University of Chicago Press, Chicago, 1996).
  
\bibitem{KapGal06}
  J.~I.~Kapusta and C.~Gale,
  Finite-Temperature Field Theory: Principles and Applications
  (Cambridge University Press, Cambridge, 2006), 2nd ed.

\bibitem{PitLif81}
  L.~P.~Pitaevskii and E.~M.~Lifshitz,
  Physical Kinetics
  (Pergamon Press, Oxford, 1981).

\end{thebibliography}
\end{document}